\documentstyle[aps,preprint]{revtex}
\begin{document}
\draft

\title{Choptuik scaling and the scale invariance of Einstein's equation}

\author{David Garfinkle
\thanks {Email: garfinkl@vela.acs.oakland.edu}}
\address{
\centerline{Department of Physics, Oakland University,
Rochester, MI 48309}}

\maketitle

\null\vspace{-1.75mm}

\begin{abstract}
The relationship of Choptuik scaling to the scale invariance of Einstein's
equation is explored.  Ordinary dynamical systems often have limit cycles:
periodic orbits that are the asymptotic limit of generic solutions.  We show
how to separate Einstein's equation into the dynamics of the overall scale and
the dynamics of the ``scale invariant'' part of the metric.  Periodicity of
the scale invariant part implies periodic self-similarity of the spacetime.
We also analyze a toy model that exhibits many of the features of Choptuik
scaling. 
\end{abstract}


\section{Introduction}

Recently Choptuik has found scaling phenomena in gravitational 
collapse.\cite{choptuik}
He numerically evolves a one parameter family of initial data for a spherically
symmetric scalar field coupled to gravity.  Some of the data collapse to form
black holes while others do not.  There is a critical value of the parameter
separating those data that form black holes from those that do not.  The
critical solution (the one corresponding to the critical parameter) has the
property of periodic self similarity: after a certain amount of logarithmic
time the profile of the scalar field repeats itself with its spatial scale
shrunk.  For parameters slightly above the critical parameter the mass of
the black hole formed scales like $ {(p-p*)}^\gamma$ where $p$ is the parameter, $p*$ is its critical value and $\gamma$ is a universal scaling
exponent that does not depend on which family is being evolved.  
Numerical simulations of the critical gravitational collapse of other types of
spherically symmetric matter were subsequently performed.  
These include complex scalar fields,\cite{eardley} perfect fluids,\cite{evans1}
axions and dilatons\cite{stewart} and Yang-Mills fields.\cite{bizon}  
In addition scaling has been found in the collapse of axisymmetric gravity 
waves.\cite{evans2}  Thus scaling
seems to be a generic feature of critical gravitational collapse.
In some of these systems the critical solution has periodic self-similarity
while in other systems it has exact self-similarity.  

These phenomena were discovered numerically, so one would like to have an
analytic explanation for why systems that just barely undergo gravitational
collapse behave in this way.  Gundlach\cite{gundlach1} and Koike, Hara 
and Adachi\cite{hara} have 
explained the scaling
of black hole mass analytically subject to the following assumptions:
(i) the critical solution is periodically self-similar and (ii) the 
critical solution has exactly one unstable mode.  This still leaves 
unexplained the mystery of why the critical solution is periodically
self-similar.

While periodic self-similarity is an unusual property for a dynamical
system, periodicity is not.  In fact, many dynamical systems have limit
cycles, {\it i.e.} periodic trajectories that are approached asymptotically
for a large class of initial conditions.  What is it about Einstein's 
equation that gives rise to periodically self-similar solutions rather
than periodic ones?  One feature of Einstein's equation is scale invariance:
let $ g_{ab}$ be a solution of the vacuum Einstein equation and let $k$
be a positive constant.  Then $ k {g_{ab}} $ is a solution of the vacuum
Einstein equation.  The same property holds for the Einstein-scalar equation
(the system studied by Choptuik).  Let $ ({g_{ab}},\phi)$ be a solution of
the Einstein-scalar equation and let $k$ be a positive constant.  Then
$ (k {g_{ab}},\phi)$ is a solution of the Einstein-scalar equation.  This
feature of scale invariance suggests that in some sense the metric can be
decomposed into an ``overall scale'' and a ``scale invariant part'' and that
these two pieces have very different dynamics.  In particular a periodically
self-similar metric could be realized as a periodic scale invariant part of
the metric.  Also an exactly self-similar metric could be realized as a
static scale invariant part of the metric.  Thus if the dynamical system of
the scale invariant part of the metric has a limit point {\it i.e.} a
point in phase space that is approached asymptotically for a large class
of initial conditions (a usual feature for dynamical systems) then the 
metric has a critical solution that is exactly self-similar.

This program of finding scale invariant dynamics for the metric 
and matter fields has been carried out in the spherically symmetric 
case for various
kinds of matter.  In spherical symmetry the metric takes the form\cite{choptuik}
$$
d {s^2} = - \, {\alpha ^2} (r,t) \, d {t^2} \; + \; {a^2} (r,t) \, d {r^2} \; + \; {r^2} \, d {\Omega ^2} \; \; \; .
\eqno(1)
$$
With $t$ or $r$ as the overall scale, the quantities $\alpha $ and $a$ 
are scale invariant.  The dynamics then becomes a set of 
differential equations for
$\alpha , \, a $ and the scale invariant matter variables.  These equations 
have been found for radiation\cite{hara}, a massless scalar 
field\cite{gundlach1}, scalar electrodynamics\cite{gundlach2} and the 
Einstein-Yang-Mills system.\cite{gundlach3}

Homogeneous cosmologies have also been treated using this sort of approach.
Wainwright and Hsu\cite{wain} have found a set of scale invariant variables 
that describe the dynamics of homogeneous, anisotropic spacetimes.  These 
variables have been useful both in the work of reference\cite{wain} and 
in recent work of Rendall.\cite{rendall}  The connection between scale 
invariance
and discrete self similarity has been noted in a study by 
Traschen\cite{jennie} of fluctuations about an extremal black hole.

One would like to carry out this program of finding scale invariant metric 
variables in cases where there are no symmetries.  The renormalization
group framework of reference\cite{hara} is sufficiently general to 
accomodate the case of scale invariance with no symmetries.  However,
the coordinate invariance of general relativity allows more than one 
possible set of scale invariant variables.  One would like to find a
set appropriate for critical graviational collapse.
In the Euclidean approach 
to quantum gravity\cite{hawking} the metric is decomposed into 
a conformal factor and a determinant=1 piece.  However, in the Lorentzian 
case Einstein's equations are a dynamical system described by the ADM 
formalism.  It therefore makes sense to have the overall scale determined 
by some dynamical condition.

In section 2 we treat the dynamics of a toy model that has many of the
features of Choptuik scaling.  In section 3 we modify the ADM formalism
to write Einstein's equation as a dynamical system of scale invariant
quantities.  Section 4 is a discussion of the implications of these
results.

\section{Toy model}

We now consider a toy model that has many of the features of Choptuik scaling.
This model is constructed by adding an overall scale degree of freedom to
a model known to have a limit cycle.
The model is a dynamical system consisting of three functions of time: $ a , b$ and $c$.  Define the quantity $s$ by
$$
s \equiv {{2 {\dot a}} \over a} \; - \; \left ( {{\dot b} \over b} \; + \;
{{\dot c} \over c} \right )
\eqno(2)
$$
where an overdot denotes derivative with respect to $t$.
Choose the equations of motion of this system to be
$$
3 \; {d \over {dt}} \; \left ( {{\dot a} \over a} \right ) = - \, \ln \left ( {{a^2} \over {bc}} \right ) \; + \; s \; + \; \epsilon \, s \, \left ( 1 \; - \; {{\left [ {1 \over 3} \; \ln \left ( {{a^2} \over {bc}} \right ) \right ] }^2} \right ) \; \; \; ,
\eqno(3)
$$
$$
3 \; {d \over {dt}} \; \left ( {{\dot b} \over b} \right ) =  \ln \left ( {{b^2} \over {ac}} \right ) \; + \; s  \; \; \; ,
\eqno(4)
$$
$$
3 \; {d \over {dt}} \; \left ( {{\dot c} \over c} \right ) = 3 \, \ln \left ( {a \over b} \right ) \; + \; s \; - \;  \epsilon \, s \, \left ( 1 \; - \; {{\left [ {1 \over 3} \; \ln \left ( {{a^2} \over {bc}} \right ) \right ] }^2} \right ) \; \; \; .
\eqno(5)
$$
Here $\epsilon $ is a small positive constant.
Note that this system has the property of scale invariance: if ($a,b,c$) is
any solutions of the equations of motion and $k$ is any positive constant then
($ka,kb,kc$) is also a solution of the equations of motion.  This suggests that
one might better understand this system by dividing its variables into an
``overall scale'' and a ``scale invariant part.''  To make this idea precise
define the quantities $ N , \, \alpha $ and $\beta$ by
$$
N \equiv {{(abc)}^{1/3}} \; \; \; ,
\eqno(6)
$$
$$
\alpha \equiv \ln ( a/N ) \; \; \; ,
\eqno(7)
$$
$$
\beta \equiv \ln ( b/N ) \; \; \; .
\eqno(8)
$$
Note that $\alpha $ and $\beta$ are scale invariant.  Thus the quantity $N$ is
the overall scale of the system and ($\alpha , \beta $) is the scale invariant
part of the system.  Expressing equations (3-5) in terms of these new variables
we find
$$
{\ddot \alpha } \; + \; {d \over {dt}} \; \left ( {{\dot N} \over N} \right ) = - \, \alpha \; + \; {\dot \alpha } \; + \; \epsilon \, {\dot \alpha } \, \left ( 1 \, - \, {\alpha ^2} \right ) \; \; \; ,
\eqno(9)
$$
$$
{\ddot \beta } \; + \; {d \over {dt}} \; \left ( {{\dot N} \over N} \right ) = \beta \; + \; {\dot \alpha }  \; \; \; ,
\eqno(10)
$$
$$
- \, {\ddot \alpha } \; - \; {\ddot \beta} \; + \; {d \over {dt}} \; \left ( {{\dot N} \over N} \right ) =  \alpha \; - \; \beta \; + \; {\dot \alpha } \; - \; \epsilon \, {\dot \alpha } \, \left ( 1 \, - \, {\alpha ^2} \right ) \; \; \; ,
\eqno(11)
$$
Adding equations (9-11) we find
$$
{d \over {dt}} \; \left ( {{\dot N} \over N} \right ) = {\dot \alpha } \; \; \; .
\eqno(12)
$$
Using this result in equations (9-10) we find
$$
{\ddot \alpha } = - \, \alpha \; + \; \epsilon \, {\dot \alpha } \, \left ( 1 \, - \, {\alpha ^2} \right ) \; \; \; ,
\eqno(13)
$$
$$
{\ddot \beta } = \beta \; \; \; .
\eqno(14)
$$
Note that the system ($\alpha ,\beta$) is a dynamical system independent of the
overall scale $N$ and with equations of motion given by equations (13-14).  The general solution of equation (14) is
$$
\beta = {1 \over 2} \; \left ( {\beta _0} \, + \, {v_{\beta 0}} \right ) \, {e^t} \; + \; {1 \over 2} \;  \left ( {\beta _0} \, - \, {v_{\beta 0}} 
\right ) \, {e^{- \, t}}
\eqno(15)
$$
where ${\beta _0} \equiv \beta (0) $ and $ {v_{\beta 0}} \equiv {\dot \beta } (0) $.  Note that as $ t \to \infty $ we have $ \beta \to \pm \infty $ or $0$
depending on whether the quantity $ {\beta _0} \, + \, {v_{\beta 0}}$ is
respectively positive, negative or zero.  Now consider a generic one parameter family of initial data for the system ($a,b,c$) depending on parameter 
$p$.  In general there will be a range of $p$ for which the long time 
evolution of the system gives $ \beta \to \infty $, a range of $p$ for which
$\beta \to - \infty $ and a critical value $p*$ of p for which $ \beta \to 0$.
This is analogous to the case of Choptuik scaling where there is a range of $p$
for which the late time evolution gives a black hole, a range of $p$ for which
the late time evolution gives flat space and a critical parameter $p*$ 
separating these two ranges for which the late time evolution gives the
Choptuik critical solution.

Now consider the evolution equation (13) for $ \alpha $.  This is the {\it van der Pol} equation.\cite{arnold}  It is well known that this equation has a 
stable limit cycle.  A solution of equation (13) to zeroth order in $\epsilon $ is
$$ 
\alpha = 2 \, \cos \left ( t \, + \, {\phi _0} \right )
\eqno(16)
$$
where $\phi _0$ is a constant.  Furthermore any initial data sufficiently
close to the trajectory given in equation (16) will approach that trajectory
asymptotically as $ t \to \infty $.  It then follows that for generic
one parameter families of initial data for the ($\alpha , \beta $) system 
(sufficiently close to the limit cycle) there is a critical value of
the parameter for which the trajectory is asymptotically periodic.

Now consider the behavior, at late times, of the overall 
scale $N$ in the critical solution.
With $\alpha $ given by equation (16) we find using equation (12)
$$
N = n \, \exp \left [ \kappa t \, + \, 2 \sin (t + {\phi _0}) \right ]
\eqno(17)
$$
where $n$ and $\kappa $ are constants.  That is, the overall scale is a
periodic function multiplied by an exponential.  The asymptotic behavior of
the critical solution expressed in terms of the variables ($a,b,c$) is then
$$
a = n \, \exp \left [ \kappa t \, + \, 2 {\sqrt 2} \, \sin \left ( t \, + \, {\phi _0} \, + \, \pi / 4 \right ) \right ] \; \; \; ,
\eqno(18)
$$
$$
b = n \, \exp \left [ \kappa t \, + \, 2 \, \sin \left ( t \, + \, {\phi _0} \right ) \right ] \; \; \; ,
\eqno(19)
$$
$$
c = n \, \exp \left [ \kappa t \, + \, 2 {\sqrt 2} \, \sin \left ( t \, + \, {\phi _0} \, - \, \pi / 4 \right ) \right ] \; \; \; .
\eqno(20)
$$
Thus the critical solution for ($a,b,c$) is periodically self similar.  After
a certain amount of time the solution repeats its behavior with only its overall scale changed.

\section{Separation of Einstein's equation}

We now apply the ideas developed in the previous section to Einstein's
equation.  Start with the standard ADM formalism with, 
for simplicity, zero shift.
The spacetime metric has the form
$$
d {s^2} = - \, {N^2} \, d {t^2} \; + \; {h_{ik}} \, d {x^i} \, d {x^k} \; \; \; .
\eqno(21)
$$
Here $h_{ik}$ is the intrinsic metric of the $t={\rm const.}$ slices.  The 
lapse $N$ can be chosen arbitrarily.  The extrinsic curvature of the
$t={\rm const.}$ slices is
$$
{K_{ik}} = {1 \over {2 N}} \; {\partial _t} \, {h_{ik}} \; \; \; .
\eqno(22)
$$
(Here $ {\partial _t} $ denotes $ \partial / \partial t$).  We would like
to keep the same time and space coordinates ($t,{x^i}$) when the 
spacetime metric is multiplied by an
overall constant.  Thus under the transformation $ {h_{ik}} \to k {h_{ik}} $
for a constant $k$ we want $ N \to {\sqrt k} \, N$.  An evolution equation
consistent with this condition is 
$$
{\partial _t} N = {1 \over 3} \; {N^2} \, K \; \; \; .
\eqno(23)
$$
Actually any constant could be chosen to replace the $1/3$; but as we will
see later the choice of $1/3$ has other nice features.  We now wish to 
replace $h_{ik}$ with a scale invariant quantity.  Define
$$
{{\tilde h}_{ik}} \equiv {N^{- 2}} \, {h_{ik}} \; \; \; .
\eqno(24)
$$
This ${\tilde h}_{ik}$ is the ``scale invariant part'' of the spatial metric.
Using equations (22) and (23) we find
$$
{\partial _t} {{\tilde h}_{ik}} = 2 \, {{\tilde K}_{ik}}
\eqno(25)
$$
where $ {\tilde K}_{ik}$, the ``scale invariant part'' of the extrinsic
curvature, is defined by
$$
{{\tilde K}_{ik}} = {N^{ - 1}} \, \left ( {K_{ik}} \; - \; {1 \over 3} \; K \, {h_{ik}} \right ) \; \; \; .
\eqno(26)
$$
Note that $ {\tilde K}_{ik}$ is trace free.  This is due to the presence of the
factor $1/3$ in equation (23).  
Thus the extrinsic curvature essentially has two
parts: the trace and the scale invariant part.
We now need to find an evolution equation for ${\tilde K}_{ik}$ and we would like to express this equation in terms of scale invariant quantities.  One
such quantity is
$$
{\omega _i} \equiv {N^{-1}} \, {\partial _i} N \; \; \; .
\eqno(27)
$$
The evolution equation for $K_{ik}$ is
$$
{\partial _t} \, {K_{ik}} = {D_i} {D_k} N \; + \; N \, \left ( 2 \, {K_{ip}} {{K_k}^p} \; - \; K \, {K_{ik}} \; + \; {R_{ik}} \; - \; {^{(3)}}{R_{ik}} \right ) \; \; \; .
\eqno(28)
$$
Here $D_i$ and ${^{(3)}}{R_{ik}}$ are respectively covariant derivative
and Ricci tensor of the metric $h_{ik}$ and $R_{ik}$ is the spacetime
Ricci tensor.  From equations (22),(23) and (28) straightforward but tedious 
algebra gives an evolution equation for ${\tilde K}_{ik}$.  Expressing
that equation in terms of scale invariant quantities yields
$$
{\partial _t} {{\tilde K}_{ik}} = - \; {2 \over 3} \; (NK) \, {{\tilde K}_{ik}}
\; + \; 2 \, {{\tilde K}_{ip}} {{{\tilde K}_k}^{\; \; p}} \; + \; {R_{ik}} \; - \; {^{(3)}}{{\tilde R}_{ik}} \; + \; 2 {{\tilde D}_i} {\omega _k} \; - \; 2 {\omega _i} {\omega _k}
$$
$$
+ \; {1 \over 3} \; {{\tilde h}_{ik}} \, \left [ {^{(3)}}{\tilde R} \; + \; 2 {\omega _p} {\omega ^p} \; - \; 2 {{\tilde D}_p} {\omega ^p} \; - \; {{\tilde h}^{pq}} {R_{pq}} \right ] \; \; \; .
\eqno(29)
$$
Here ${\tilde D}_i$ and ${^{(3)}}{{\tilde R}_{ik}}$ are 
respectively covariant derivative
and Ricci tensor of the metric ${\tilde h}_{ik}$ and all indices are 
lowered and raised with $ {\tilde h}_{ik}$ and its inverse.  

We now extract from Einstein's equation the equations of a dynamical system 
whose variables are scale invariant quantities.  For the vacuum case these
variables are (${{\tilde h}_{ik}}, \, {{\tilde K}_{ik}}, \, {\omega _k}$).
In the case where matter fields are present, and where the matter field
equations are scale invariant, these metric variables must be supplemented
with scale invariant matter fields.  Is the set (${{\tilde h}_{ik}}, \, {{\tilde K}_{ik}}, \, {\omega _k}$), along with the matter fields, complete?
That is, can the time derivative of each variable be expressed in terms of
the other variables?  Clearly this holds for the time derivative of 
$ {\tilde h}_{ik}$.  In equation (29) the only questionable term is the one
proportional to $N K$.  However, we now show that the Hamiltonian constraint
gives $N K$ in terms of (${{\tilde h}_{ik}}, \, {{\tilde K}_{ik}}, \, {\omega _k}$) and the matter fields.  The Hamiltonian constraint is 
$$
{^{(3)}}R \; + \; {K^2} \; - \; {K_{ik}} {K^{ik}} = 2 \, {G_{\mu \nu}} {n^\mu} {n^\nu} \; \; \; .
\eqno(30)
$$
Expressing this constraint in terms of scale invariant quantities we have
$$
N K = \pm \; {\sqrt {3 \over 2}} \; {{\left [ {{\tilde K}_{pq}} {{\tilde K}^{pq}} \; + \; 2 \, {\omega _p} {\omega ^p} \; + \; 4 {{\tilde D}_p} {\omega ^p} \; - \; {^{(3)}}{\tilde R} \; + \; 2 {{\tilde h}^{pq}} {R_{pq}} \; - \; {N^2} \, R \right ] }^{1/2}} \; \; \; .
\eqno(31)
$$
It is also helpful to have an evolution equation for $ N K$.  From equations
(22),(23) and (28) it follows that
$$ 
{\partial _t} ( N K ) = - \; {2 \over 3} \; {{(N K)}^2} \; + \; 5 \, {{\tilde D}_p} {\omega ^p} \; + \; 4 {\omega _p} {\omega ^p} \; + \; {{\tilde h}^{pq}} {R_{pq}} \; - \; {^{(3)}}{\tilde R} \; \; \; .
\eqno(32)
$$
In equations (31) and (32) all indices are lowered and raised with ${\tilde h}_{ik}$ and its inverse.  We also need an evolution equation for $ \omega _k$.
From equations (23) and (27) it follows that
$$
{\partial _t} {\omega _k} = {1 \over 3} \; {{\tilde D}_k} \, (N K) \; \; \; .
\eqno(33)
$$
Equations (25), (29) and (33) are the evolution equations 
for the dynamical system
of scale invariant quantities (${{\tilde h}_{ik}}, \, {{\tilde K}_{ik}}, \, {\omega _k}$).  Here the auxiliary quantity $ N K$ is given by the constraint
equation (31) and its evolution is given by equation (32).  

Suppose that we have a solution for the dynamical system (${{\tilde h}_{ik}}, \, {{\tilde K}_{ik}}, \, {\omega _k}$).  What else is needed to determine the
metric?  Clearly the quantity $N$ along with $ {\tilde h}_{ik}$ is enough to
determine the spacetime metric.  However, the quantity $ \omega _k$ already
contains most of the information about $N$.  The only missing piece of
information is the value of $N$ at a single point of space as a function of
time.  Pick a spatial point $x^i_0$ and define $ {N_0}(t) \equiv N(t, {x^i_0})$.  Then the quantity $N_0$ along with a solution of the scale
invariant dynamical system determines the spacetime metric.  How does the
quantity $N_0$ evolve?  Applying equation (23) at the point $ x^i_0$ we find
$$
{\partial _t} \ln {N_0} = {1 \over 3} (N K) ({x^i_0}) \; \; \; .
\eqno(34)
$$
However, the quantity $N K$ is determined by the dynamical system (${{\tilde h}_{ik}}, \, {{\tilde K}_{ik}}, \, {\omega _k}$).  Therefore, up to an overall
constant scale (the value of $N_0$ at some initial time $t_0$) the spacetime
metric is determined by the scale invariant dynamical system.  

Now suppose that the variables (${{\tilde h}_{ik}}, \, {{\tilde K}_{ik}}, \, {\omega _k}$) are periodic functions of time.  Then it follows that $N_0$ is
an exponential function multiplied by a periodic function.  It then follows that the spacetime is periodically self-similar.  If instead  
(${{\tilde h}_{ik}}, \, {{\tilde K}_{ik}}, \, {\omega _k}$) are independent
of time then $N_0$ is an exponential function of time and the spacetime
is exactly self-similar.

We now consider how to specify the scale invariant equations for one particular
type of matter: the massless, minimally coupled scalar field.  Here the Ricci
tensor is
$$
{R_{\mu \nu}} = 8 \pi {\nabla _\mu} \phi {\nabla _\nu} \phi 
\eqno(35)
$$
and the scalar field satisfies the wave equation
$$
{\nabla _\mu } {\nabla ^\mu} \phi = 0 \; \; \; .
\eqno(36)
$$
The field $\phi$ is scale invariant, so the full set of scale invariant 
quantities is (${{\tilde h}_{ik}}, \, {{\tilde K}_{ik}}, \, {\omega _k}, \, \phi$).  It follows that 
$$
{R_{ik}} = 8 \pi \, {{\tilde D}_i} \phi {{\tilde D}_k} \phi \; \; \; ,
\eqno(37)
$$
$$
{N^2} \, R = 8 \pi \left [ {{\tilde D}_i} \phi {{\tilde D}^i} \phi \; - \; {{( {\partial _t} \phi )}^2} \right ] \; \; \; .
\eqno(38)
$$
Expressed in terms of scale invariant quantities equation (36) becomes
$$
- \, {\partial _t} {\partial _t} \phi \; - \; {2 \over 3} \; N K \, {\partial _t} \phi \; + \; 2 {\omega ^i} {{\tilde D}_i} \phi \; + \; {{\tilde D}_i} {{\tilde D}^i} \phi = 0 \; \; \; .
\eqno(39)
$$

\section{Discussion}

The analysis of the previous section shows that the Einstein-scalar equation
can be separated into equations for scale invariant quantities and an
equation for the overall scale.  Periodic self similarity of the Choptuik
critical solution is then equivalent to the existence of a limit cycle for
the scale invariant system.  Thus the somewhat odd property of periodic
self similarity is reduced to the more familiar property of limit cycles
of dynamical systems.  (Correspondingly the property of exact self-similarity
is reduced to the property of limit points of dynamical systems).

What remains to be seen is whether the particular set of scale invariant 
variables used in section 3 is appropriate for the Einstein-scalar system.  
That is, we need to find out whether the variables 
(${{\tilde h}_{ik}}, \, {{\tilde K}_{ik}}, \, {\omega _k}, \, \phi$) are 
periodic functions of time in the Choptuik critical solution.  
This question is presently under study.  If the answer is no, then one would 
need to search for a different set of scale invariant variables to 
describe critical gravitational collapse.

One would also like to know what physical processes are responsible for the existence of limit cycles in the process of gravitational collapse.
In the {\it van der Pol} equation,
limit cycles come about due to energy dissipation (both positive and negative).
The {\it van der Pol} system is essentially the harmonic oscillator with a
small perturbation.
For the harmonic oscillator energy $ E= ( {{\dot \alpha }^2} + {\alpha ^2})/2$
one can show using equation (13) that on the average 
the energy increases if the
amplitude of oscillations is below that of the limit cycle and decreases if the
amplitude is above that of the limit cycle.  Thus the process of energy 
dissipation drives arbitrary trajectories to the limit cycle.  What is the
analog of energy dissipation (both positive and negative) in the spherically
symmetric Einstein-scalar system?  When the scalar field is weakly gravitating
it disperses at late times; so the energy in a fixed region tends to decrease.
For a strongly self gravitating scalar field the self gravity 
tends to concentrate 
the energy of the field into ever smaller regions.  It is
these two competing effects that, depending on their relative strengths,
combine to form field dispersion, black hole formation or the critical
solution.  What is needed is to find an ``energy-like'' quantity for
the scale invariant system whose evolution corresponds to effects of 
concentration or dispersion of the field.  

What about the more general case of systems with axisymmetry or no symmetry at
all?  Here the equations become much more complicated and proving the 
existence of limit cycles becomes much more difficult.  Here too a
promising approach is to find an energy like quantity that is conserved 
on the limit cycle.  An examination of the behavior of such a quantity
should yield new intuitions on how scaling arises.  It is therefore likely
that an examination of the scale invariant dynamical system will
be a powerful tool in understanding Choptuik scaling.

\section{Acknowledgements}

It is a pleasure to thank Beverly K. Berger, Carsten Gundlach and Alan 
Rendall for helpful discussions.
This work was partially supported by a Cottrell College Science Award of 
Research Corporation to Oakland University.


\begin{references}

\bibitem{choptuik}
M. Choptuik, Phys. Rev. Lett. {\bf 70}, 9  (1993)

\bibitem{eardley}
E.W. Hirschmann and D.M. Eardley, Phys. Rev. {\bf D51}, 4198 (1995)

\bibitem{evans1}
C.R. Evans and J.S. Coleman, Phys. Rev. Lett. {\bf 72}, 1782 (1994)

\bibitem{stewart}
R.S. Hamade, J.H. Horne and J.M. Stewart, gr-qc/9511024

\bibitem{bizon}
M. Choptuik, T. Chmaj and P. Bizon, Phys. Rev. Lett. {\bf 77}, 424 (1996)

\bibitem{evans2}
A.M. Abrahams and C.R. Evans, Phys. Rev. Lett. {\bf 70}, 2980 (1993)

\bibitem{gundlach1}
C. Gundlach, Phys. Rev. {\bf D55}, 695 (1997)

\bibitem{hara}
T. Koike, T. Hara and S. Adachi, Phys. Rev. Lett. {\bf 74}, 5170 (1995),
T. Hara, T. Koike and S. Adachi, gr-qc/9607010 

\bibitem{gundlach2}
C. Gundlach and J. Martin-Garcia, Phys. Rev. {\bf D54}, 7353 (1996)

\bibitem{gundlach3}
C. Gundlach, gr-qc/9610069

\bibitem{wain}
J. Wainwright and L. Hsu, Class. Quantum Grav. {\bf 6}, 1409 (1989)

\bibitem{rendall}
A. Rendall, gr-qc/9703036

\bibitem{jennie}
J. Traschen, Phys. Rev. {\bf D50}, 7144 (1994)

\bibitem{hawking}
S.W. Hawking, in {\it General Relativity, an Einstein Centenary Survey}, ed. S. W. Hawking and W. Israel, Cambridge University Press (Cambridge, 1979)

\bibitem{arnold}
{\it Ordinary differential equations} by V.I. Arnold, The MIT press (Cambridge,
Mass. 1991)

\end{references}
\end{document}